\documentclass{rjparticle}
\usepackage{graphicx}
\usepackage{slashed}

\newcommand{\miktex}{\hbox{Mik\kern-.15em\TeX}}

\title{Status of CMS dark matter searches in 2011} 
\author{Sezen Sekmen \\ for the CMS Collaboration} 
\affil{Department of Physics, Florida State University,\\
Tallahassee, Florida 32306, USA\\Email: {\em ssekmen@cern.ch}}

\begin{document}
\maketitle
\begin{abstract}
We present the status of dark matter searches performed by the Compact Muon Solenoid Experiment using 7 TeV $pp$ data collected by the CERN Large Hadron Collider in 2010 and 2011.  
The majority of the results shown here were obtained using 1.1 fb$^{-1}$ of data.  We give highlights from analyses searching for candidates such as WIMPs, gravitinos, axinos and TeV scale particles.  All observations so far were found to be consistent with the Standard Model predictions.  The search results were used to set exclusion limits on various new physics scenarios.
\end{abstract}

\section{Introduction}
\label{sec:introduction}

The quest for new physics has entered a fascinating era with the start of the CERN Large Hadron Collider (LHC)~\cite{LHC} in 2009.  As of November 2011, over 5 fb$^{-1}$ of 7 TeV proton-proton collision data have been collected, and are being relentlessly analyzed to search for any hint of 
new physics  that may lead to the replacement of the Standard Model (SM) of particle physics with a better theory.  These searches are unique in that they are guided only by the unknowns---the puzzling deficiencies of the SM---and not by any concrete theory or experimental result that indicates where the new physics may dwell.  Some guiding questions include: what is the mechanism of electroweak symmetry breaking, what is the origin of flavor, do all the forces unify and if so, how, and (assuming this is the correct question to ask) how should one construct a quantum theory of gravity?  

Another outstanding mystery, which particle physics tries to unravel together with cosmology, is the nature of dark matter in the Universe, whose existence is suggested by the astrophysics observations.  A great variety of new physics models have been devised to address these issues, and naturally, most of these models accommodate candidates for dark matter.  

The close link between particle physics and cosmology suggests that given a clue for new physics at the LHC, we are likely to extract some information also on the nature of dark matter.  A gigantic effort is ongoing in all LHC experiments to discover such a clue.  Here, we will focus on the Compact Muon Solenoid (CMS) experiment that leads a diverse program for new physics searches, and summarize the current studies which may particularly help our understanding of dark matter.  After a short introduction to CMS, we will give the status of searches that look for a variety of dark matter candidates, ranging from weakly interacting massive particles to gravitinos and axinos to TeV-scale dark matter candidates.  Except a few cases, the results we show were obtained using $\sim$1.1 fb$^{-1}$ of LHC data.  So far, all CMS measurements are consistent with the SM.

\section{The CMS detector}

\begin{figure}[htbp]
\begin{center}
\includegraphics[height=4.7cm]{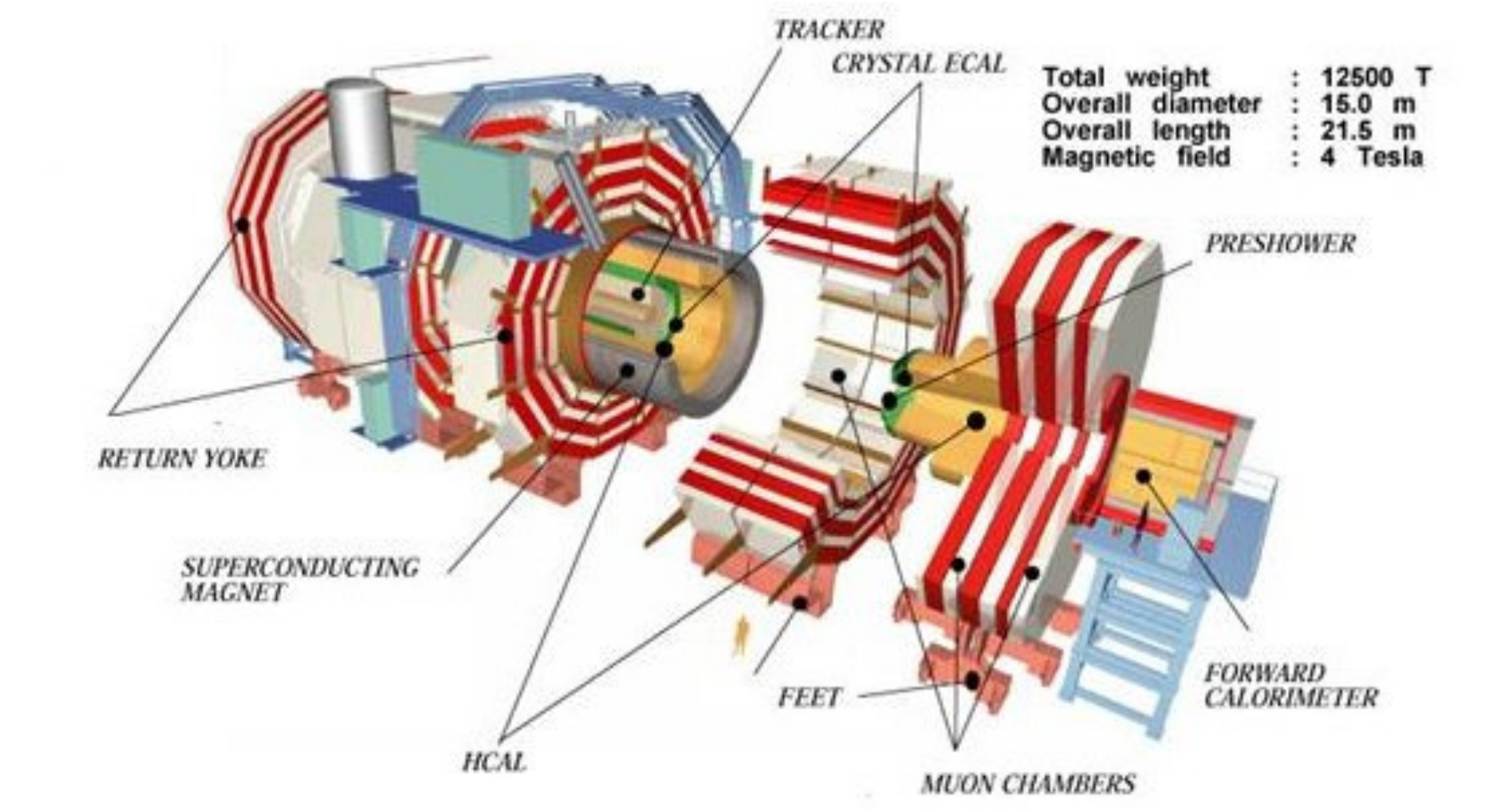}
\caption{Schematic drawing of the CMS detector}
\label{fig:cms}
\end{center}
\end{figure}

CMS is one of the two generic-purpose detectors at the LHC.  Figure~\ref{fig:cms} shows a schematic drawing of the CMS detector.  The central feature of the detector is a superconducting solenoid providing an axial magnetic field of 3.8 T. The bore of this solenoid is instrumented with several particle detection systems. Charged particle trajectories are measured within the field volume by a silicon pixel and strip tracker system, with full azimuthal ($\phi$) coverage and a pseudorapidity ($\eta$) acceptance from $-2.5$ to $+2.5$. Here, $\eta = -\ln(\tan(\theta/2))$ and $\theta$ is the polar angle with respect to the counterclockwise beam direction. The tracking volume is surrounded by a lead tungstate crystal electromagnetic calorimeter (ECAL) and a brass/scintillator hadron calorimeter that provide an $\eta$ coverage from $-3$ to $+3$. The forward hadron calorimeter extends the calorimetric coverage symmetrically to $|\eta| < 5$. Muons are identified in gas ionization detectors embedded in the steel return yoke of the magnet. The CMS detector is nearly hermetic, which allows for momentum-balance measurements in the plane transverse to the beam axis.  A more detailed description of the CMS detector can be found in~\cite{CMS}. 

\section{The CMS dark matter search}

CMS is an approximately hermetic instrument which can detect objects such as jets, $b-$jets, electrons, muons, taus and missing transverse energy ($\slashed{E}_T$) precisely, which makes it a general purpose detector capable of looking for a variety of dark matter particle candidates.  Currently, the most commonly explored candidates are the weakly interacting massive particles (WIMPs) such as neutralinos or sneutrinos suggested by supersymmetry (SUSY), Kaluza-Klein particles suggested by models with universal extra dimensions (UED), or the lightest T-odd parity particles suggested by little Higgs models.  There also exist searches for other supersymmetric candidates such as gravitinos or axinos through indirect hints from heavy stable charged particles (HCSPs).  In addition to these, other analyses look for dark photons, which may point to TeV scale dark matter particles.

The road from $pp$ collisions to any inference on the nature and properties of dark matter is long and arduous.  First, CMS has to discover a deviation from the SM in one or more final states.  Then, investigations will be made to determine if the discovered signal is due to new physics that includes a dark matter candidate.  This may also yield a rough estimate of the dark matter particle's mass.  Subsequently, the masses of all other particles related to the new physics will be measured based on the kinematics of the high branching ratio decays.  This will be followed by precision measurements involving cross sections, branching ratios, angular distributions and rare decays that may need 14 TeV of center of mass energy and at least O(10~fb$^{-1}$) of integrated luminosity to explore fully.  All these investigations will eventually lead to calculations of quantities such as the dark matter relic density and various interaction cross sections involving the dark matter particles.  

However, as noted earlier, CMS has yet to see any deviation from the SM predictions.  Therefore, the current CMS activity is mainly focused on mapping this consistency with the SM into exclusions on candidate new physics models.  So far, all such new physics interpretations that can be related to dark matter are done within the framework of supersymmetry (SUSY).  In the following, we will summarize the results of searches for new physics related to dark matter, and the interpretation of these results.  Current searches can be classified as follows: {\it i)} searches with missing energy final states; {\it ii)} searches for heavy stable charged particles and {\it iii)} searches for lepton jets.

\subsection{Searches with missing energy final states}

The most conventional way to search for many classical SUSY scenarios with WIMP dark matter candidates like neutralinos or sneutrinos is to look for an excess of events with high missing energy.  Such SUSY scenarios are typically characterized by dominant direct production of squarks and gluinos, and occasionally of charginos and neutralinos at the LHC, followed by cascade decays involving leptons and jets into a pair of heavy, neutral, stable particles.  All these lead to a diverse set of final states with missing energy, which are systematically explored by CMS.  Table~\ref{tab:metsearches} lists the missing energy final states subject to CMS SUSY searches.

\begin{table}[htdp]
\caption{CMS 2011 SUSY searches using missing energy final states}
\begin{center}
\begin{tabular}{|l|c|c|}
\hline
Final state & $\int L dt$ & Ref \\
                  & (fb$^{-1}$)  &       \\
\hline
jets $+ \slashed{E}_T$ (using the $\alpha_T$ variable~\cite{alphaT}) & 1.1 & \cite{RA1} \\
jets $+ \slashed{E}_T$ (using the razor variable~\cite{razor}) & 0.8 & \cite{RAzr} \\ 
jets $+ \slashed{E}_T$ (using the $M_{T2}$ variable~\cite{mt2}) & 1.1 & \cite{RAMT2} \\ 
jets $+$ missing transverse momentum & 1.1 & \cite{RA2} \\ 
jets $+ b-$jets $+ \slashed{E}_T$ & 1.1 & \cite{RA2b} \\
photons $+$ jets $+ \slashed{E}_T$ & 1.1 & \cite{RA3} \\
single lepton $+$ jets $+$ transverse momentum & 1.1 & \cite{RA4} \\ 
same-sign dileptons $+$ jets $+ \slashed{E}_T$ & 0.98 & \cite{RA5} \\
opposite-sign dileptons $+$ jets $+ \slashed{E}_T$ & 0.98 & \cite{RA6} \\ 
$Z + \slashed{E}_T$ & 0.98 & \cite{RAZ} \\
$Z + \slashed{E}_T +$ jets & 2.1 & \cite{RAZj1} \\
                                       & 0.191 & \cite{RAZj2} \\
multileptons $+ \slashed{E}_T$ & 2.1 & \cite{RA7} \\
\hline
\end{tabular}
\end{center}
\label{tab:metsearches}
\end{table}%

\begin{figure}[htbp]
\begin{center}
\includegraphics[height=6cm]{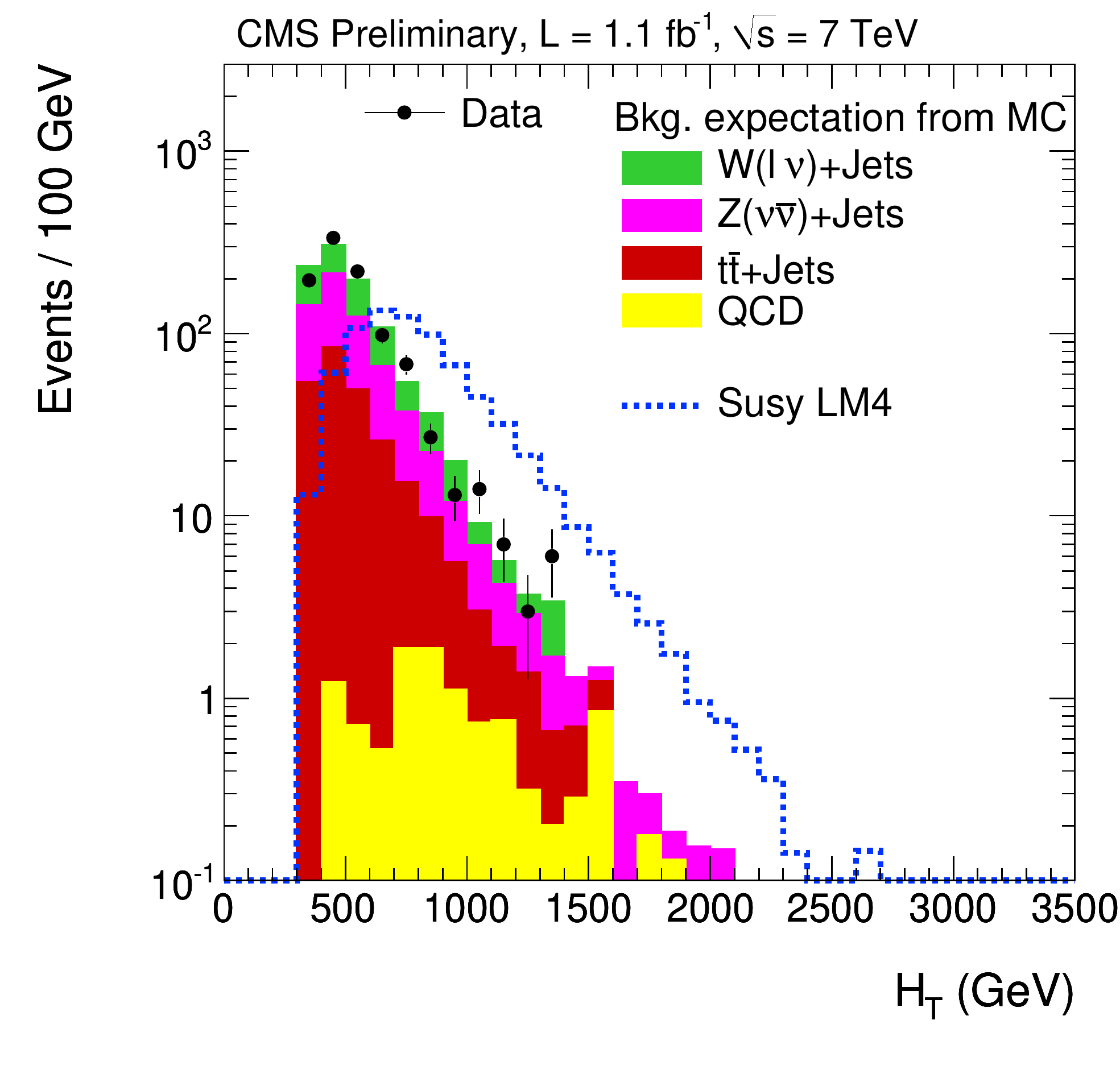} 
\includegraphics[height=6cm]{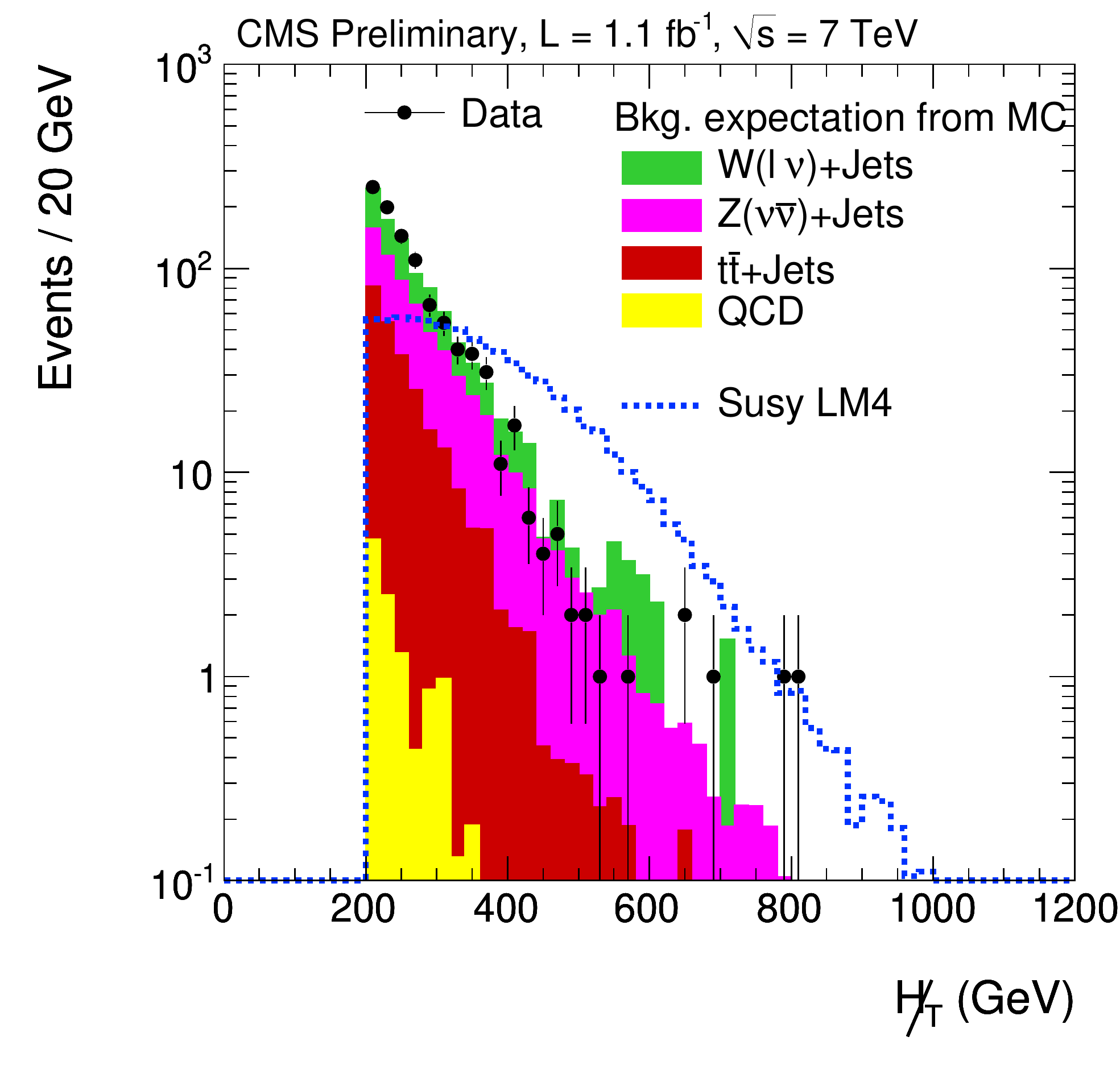} 
\caption{Distributions of hadronic transverse momentum $H_T$ (left) and missing hadronic transverse momentum $\slashed{H_T}$ (right) for data and various Monte Carlo simulation samples after all baseline selection cuts except those on $H_T$ and $\slashed{H}_T$ respectively (from the jets $+$ missing transverse momentum analysis~\cite{RA2}).}
\label{fig:RA22011_HTMHT}
\end{center}
\end{figure}

Figure~\ref{fig:RA22011_HTMHT} shows a comparison of the distributions of data with various SM and SUSY Monte Carlo simulations as a function of the hadronic transverse momentum $H_T = \sum_i p_T^{jet_i}$ (left) and missing hadronic transverse momentum $\slashed{H}_T = \sum_i \vec{p}_T^{jet_i}$ (right).  This illustrates the consistency of the LHC data with the SM predictions~\footnote{Note however that this figure is only an illustration, and that CMS analyses generally quantify SM background yields based on data-driven estimation methods.}.  Though there is no significant excess, there are some events at the high missing transverse momentum tail.  

\begin{figure}[htbp]
\begin{center}
\includegraphics[height=6cm]{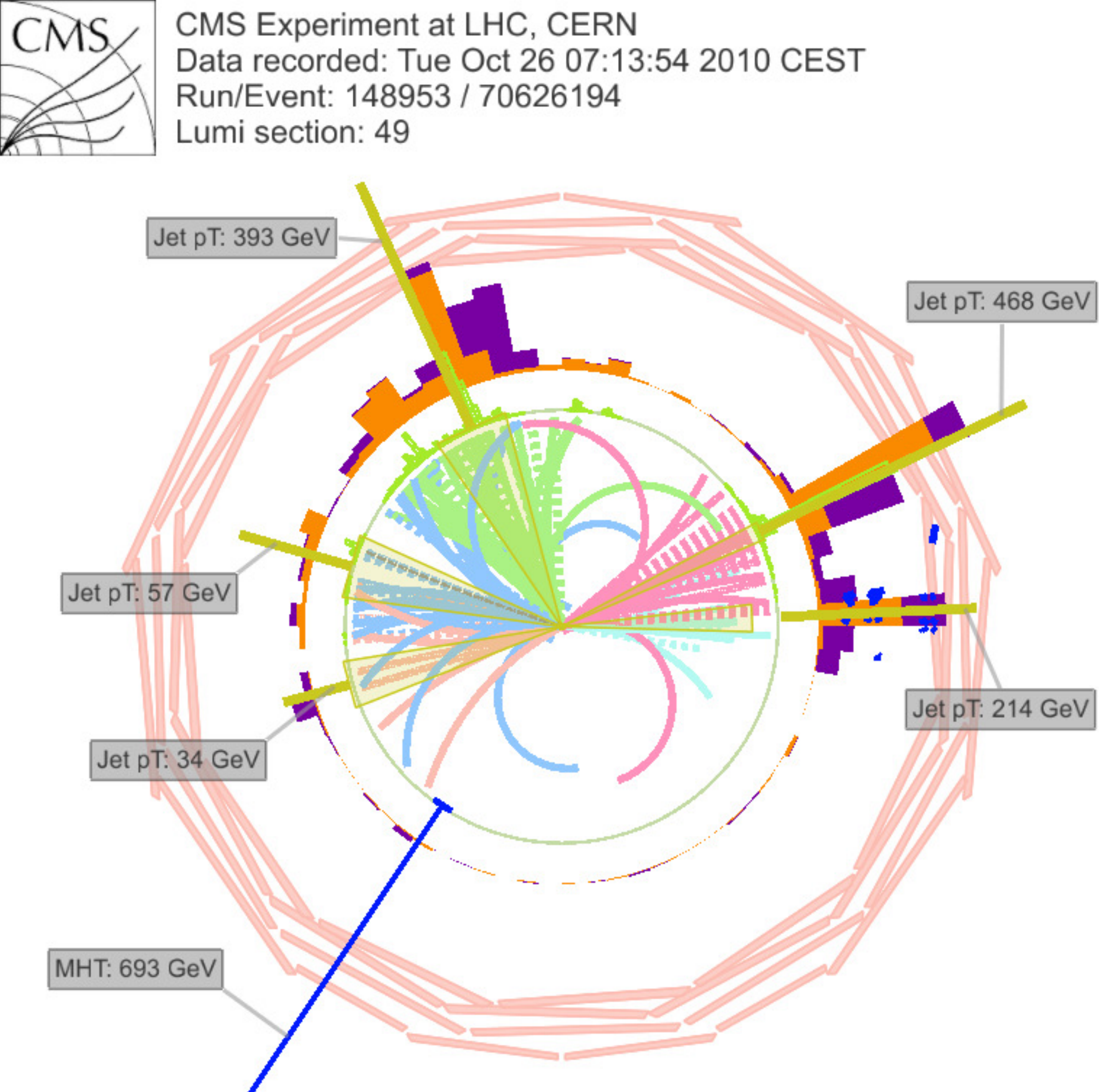}
\caption{Event 70626194, in luminosity section 49 of run 148953. ($r, \phi$) view of the highest missing transverse momentum event passing the event selection imposed in the 2010 jets $+$ missing transverse momentum analysis~\cite{RA22010}.}
\label{fig:highMHTevent}
\end{center}
\end{figure}

An event display for a high missing transverse momentum event from the 2010 dataset is shown in Figure~\ref{fig:highMHTevent} in order to illustrate the kinematics of such events.  The high missing transverse momentum in this striking event was checked to be not due to any detector effect. Furthermore, none of the jets were identified as originating from a heavy flavour quark, and none of the jet invariant mass combinations matches the $W$ or top masses.

\begin{figure}[htbp]
\begin{center}
\includegraphics[height=7.5cm]{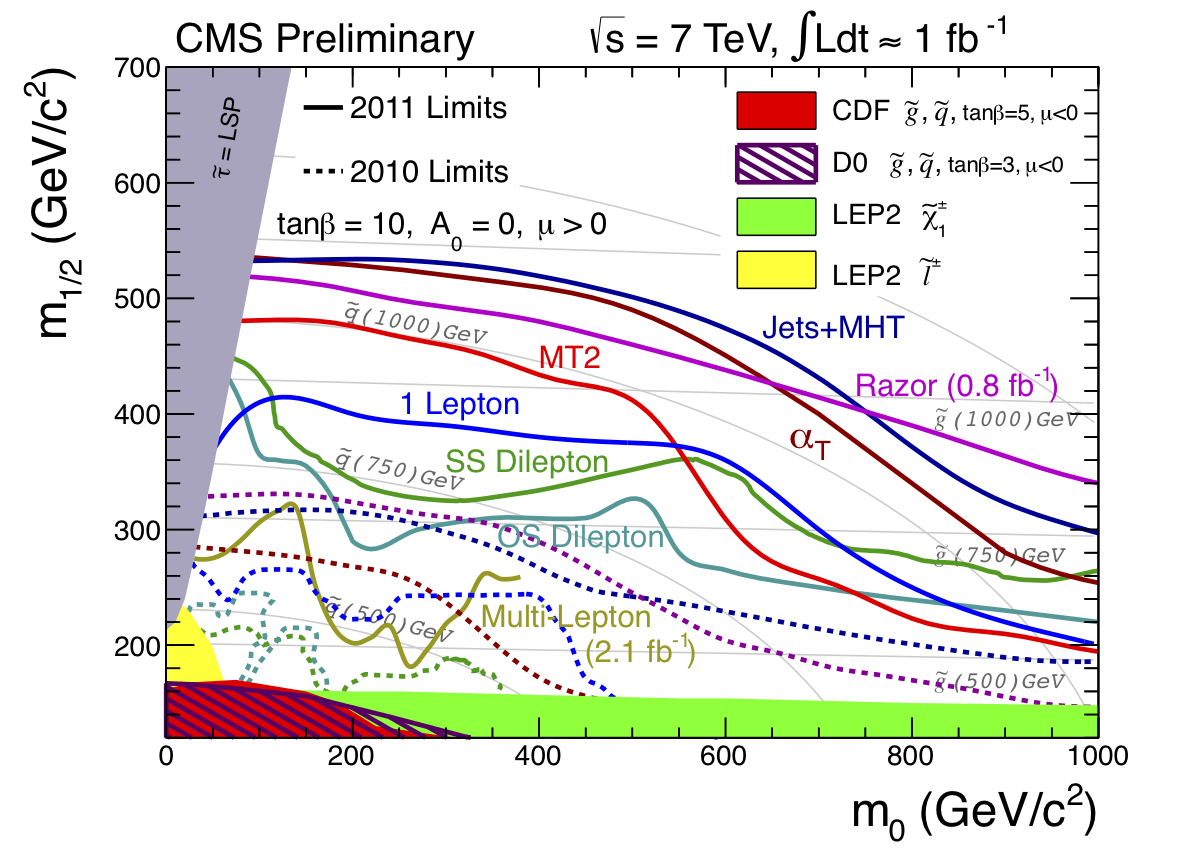}
\caption{95\%CL exclusion curves obtained by various analyses using the 2010 and 2011 data on the $m_0 - m_{1/2}$ plane with fixed $A_0 = 0$, $\tan\beta = 10$ and $\mu > 0$.}
\label{fig:mSUGRA2011}
\end{center}
\end{figure}

The most popular model that has been used for interpreting new physics searches with missing energy is the constrained MSSM (CMSSM) which is defined by four free parameters and a sign ($m_0$, $m_{1/2}$, $A_0$, $\tan\beta$ and $sgn(\mu)$) at the GUT scale.  Figure~\ref{fig:mSUGRA2011} shows the 95\% CL exclusion curves obtained by various analyses using the 2010 and 2011 data on the $m_0 - m_{1/2}$ plane with fixed $A_0 = 0$, $\tan\beta = 10$ and $\mu > 0$.  We see that, assuming neutralino dark matter, the low $m_0 - m_{1/2}$ bulk region and some parts of the stau coannihilation region (that give relic densities consistent with the WMAP measurements) are disfavored.  A large part of the favored regions predict relic densities higher than the observed WMAP upper limit, which might indicate viability of non-WIMP candidates such as axions/axinos or gravitinos.

CMS results from the analyses involving photons were interpreted within the context of general gauge mediation (GGM) models.  Here, the lightest supersymmetric particle (LSP) and hence, the dark matter candidate is a light, O($\sim$MeV) gravitino, and the next-to-lightest supersymmetric particle (NLSP) is a neutralino ($\tilde{\chi}_1^0$), which decays promptly to the gravitino plus a photon.  Interpretations were made for two cases:
\begin{itemize}
\item for a bino-like $\tilde{\chi}_1^0$ LSP, plus a gluino ($\tilde{g}$) or squark ($\tilde{q}$) next-to-next-to-lightest supersymmetric particle (NNLSP) that decays to quarks $+$ NLSP, 
\item for a purely wino $\tilde{\chi}_1^0$ LSP, plus a wino-like chargino ($\tilde{\chi}_1^\pm$) NNLSP which is almost mass degenerate with the neutralino and hence decays directly to $W + \tilde{G}$.  
\end{itemize}

\begin{figure}[htbp]
\begin{center}
\includegraphics[height=5cm]{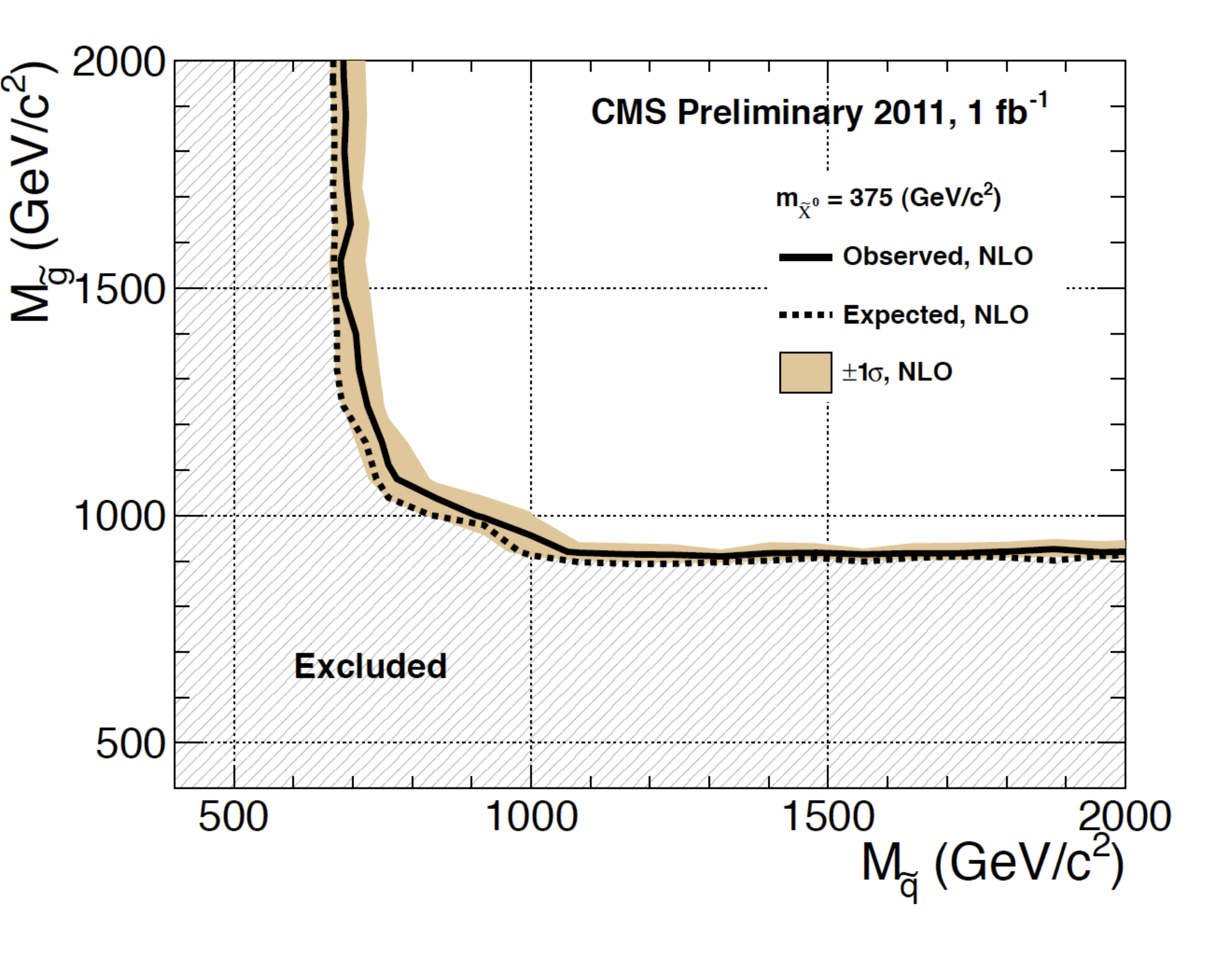}
\includegraphics[height=5.6cm]{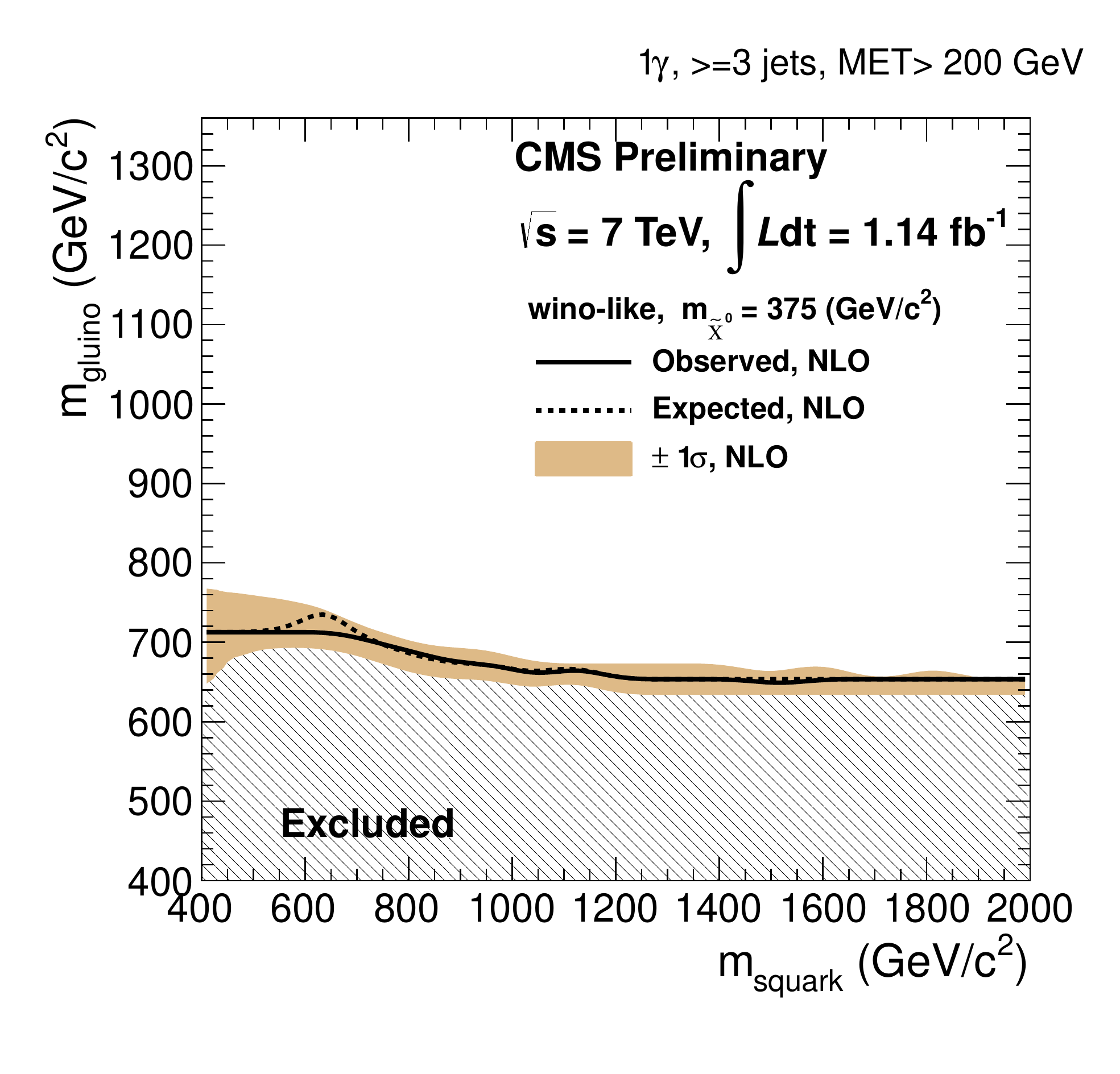}
\caption{Lower 95\% CL exclusion contours on the squark-gluino mass plane for GGM benchmark models with a fixed $\tilde{\chi}_1^0$ mass of 375 GeV.  Contours are shown for a bino-like $\tilde{\chi}_1^0$, using results from the $\ge 2$  photon $+$ jets $+ \slashed{E}_T$ search (left), and for a wino-like $\tilde{\chi}_1^0$, using results from the $= 1$ photon $+ \ge 3$ jets $+ \slashed{E}_T$ search (right).~\cite{RA3}.}
\label{fig:GGMmgmq2011}
\end{center}
\end{figure}

No signal was observed in the photon channels, and this is consistent with the constraints from Big Bang Nucleosynthesis (BBN) that disfavor SUSY models with a gravitino LSP and a neutralino NLSP.  Figure~\ref{fig:GGMmgmq2011} shows the lower 95\% CL exclusion contours on the squark-gluino mass plane for GGM benchmark models with a fixed $\tilde{\chi}_1^0$ mass of 375 GeV.  Contours are shown for a bino-like $\tilde{\chi}_1^0$, using results from the $\ge 2$  photon $+$ jets $+ \slashed{E}_T$ search (left), and for a wino-like $\tilde{\chi}_1^0$, using results from the $= 1$ photon $+ \ge 3$ jets $+ \slashed{E}_T$ search (right).

\subsection{Searches for heavy stable charged particles}

CMS has performed other more exotic searches that looked for heavy stable charged particles (HSCPs).  HSCPs are, as the name suggests, heavy and charged particles that can traverse the whole detector before they decay.  Thus, HCSPs would look like ``non-relativistic" muons in the detector, where their relatively low speed is due to their non-negligible mass.  Discovering HCSPs would suggest non-WIMP dark matter such as gravitinos and axinos, which have relatively weak couplings to other sparticles.  In cases where the gravitino or axino is the LSP and a charged slepton or squark is the NLSP, the weak gravitino coupling implies that the charged NLSP will have a long enough lifetime to traverse the detector before it decays.  

\begin{figure}[htbp]
\begin{center}
\includegraphics[height=6cm]{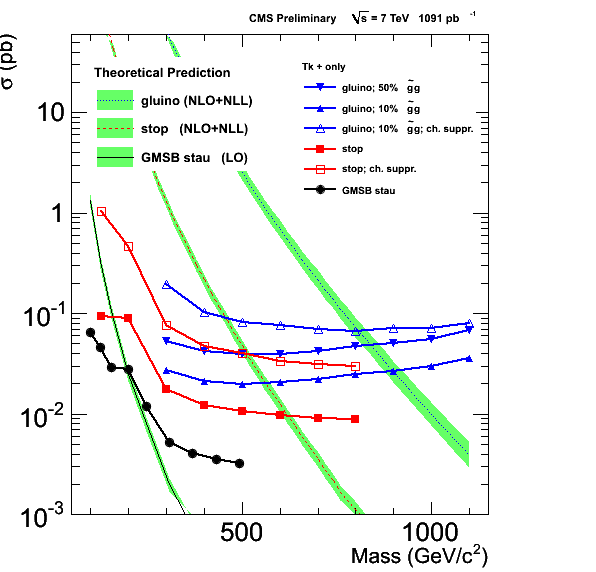}
\includegraphics[height=6cm]{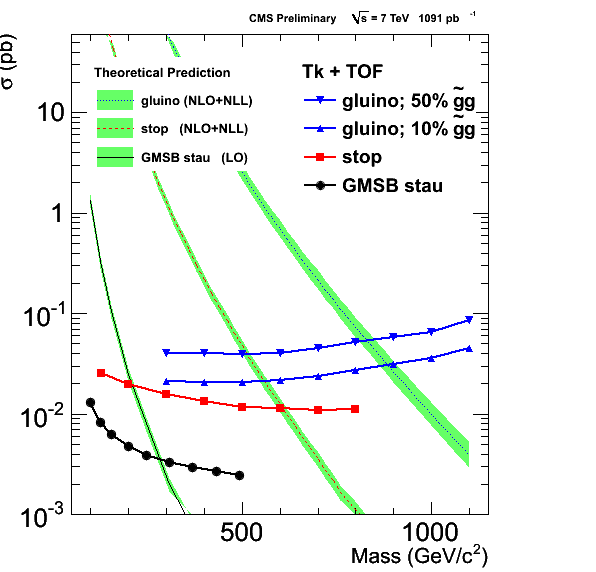}
\caption{Predicted theoretical cross sections and observed 95\% CL upper limits on the cross sections for different combinations of models and scenarios considered in the analysis.  Results are shown for the tracker-only selection (left) and tracker plus muon selection (right).
The bands represent the theoretical uncertainties on the cross section values~\cite{cmshscp}.}
\label{fig:HSCPexc2011}
\end{center}
\end{figure}

Searches for HCSPs were conducted by selecting tracks reconstructed in the inner tracker detector that have a large ionization loss ($dE/dx$) and high $p_T$.  A further selection was also studied, which additionally requires that these tracks be identified in the muon system and that they have large time of flight~\cite{cmshscp}.  Since no excess was observed, lower limits were set on the masses of various HCSP candidates.  Figure~\ref{fig:HSCPexc2011} shows predicted theoretical cross sections and observed 95\% CL upper limits on the cross section for gluinos and stops, for various nuclear interaction models.  Results are shown for 1.1 fb$^{-1}$ data, for the tracker-only selection (left) and tracker plus muon selection (right).

\subsection{Searches for lepton jets}

\begin{figure}[htbp]
\begin{center}
\includegraphics[height=6cm]{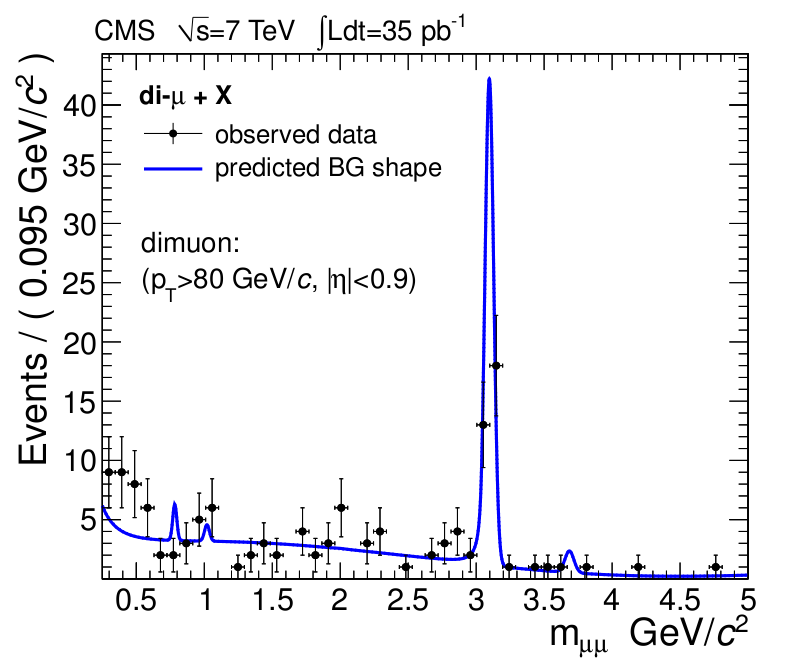}
\caption{Dimuon invariant mass distribution for events with a single dimuon of $p_T > 80$~GeV compared with the expected background. If new physics had been discovered in this channel, it would appear as a narrow peak above the background curve (with resolution similar to that of the $J/\psi$, which is the detector resolution)~\cite{cmsdarkphoton}. }
\label{fig:dimuoninvmass}
\end{center}
\end{figure}

Recent astrophysical observations showing an excess of high energy positrons in the cosmic ray spectrum~\cite{CRexcess} have inspired various new models~\cite{TeVDM} in which this excess arises from the annihilation of TeV-scale dark matter particles in the galactic halo.  Such models may also account for the observed discrepancies in direct searches for dark matter.  Some realizations of these models typically accommodate an extra $U(1)$ symmetry that couples weakly to the SM.  Breaking of this $U(1)$ results in a light vector boson with mass $\sim$O(1 GeV), called a ``hidden" or a ``dark" photon.  These dark photons have a small kinetic mixing with the SM, which allows them to decay into collimated lepton pairs, or occasionally to hadrons.  In supersymmetric extensions of such models, dark photons can be produced in the SUSY cascade decays.  CMS has performed a search using 35 pb$^{-1}$ of 2010 data for groups of collimated muon pairs---also called ``lepton jets"---to investigate the existence of dark photons or other non-SM low mass resonances. However, again, the results are consistent with the SM~\cite{cmsdarkphoton}.  Figure~\ref{fig:dimuoninvmass} shows the dimuon invariant mass distribution for events with a single dimuon with $p_T > 80$~GeV compared with the background (left), where the consistency with the SM is evident.  

\section{Conclusions}

We have presented the current status of searches by the Compact Muon Solenoid  collaboration for new physics involving dark matter,  based on analyses of 7 TeV proton-proton collision data collected in 2010 and 2011 at the LHC.  Most of the results shown here are based on 1.1 fb$^{-1}$ of data.  We gave highlights from several CMS analyses searching for dark matter candidates such as WIMPs, gravitinos, axinos and TeV scale particles.  So far, no search has observed any trace of new physics. Consequently, we have, as yet, no direct clues of the nature of dark matter.  The observed consistency with the SM has been used to set exclusion limits on various new physics models.  

The LHC is still young---it has much more data to collect, and higher collision energies to attain---thus, the quest for new physics has only just begun.  In the event of a discovery, a long series of rigorous interpretation studies will await us, which would necessarily involve the  combination of collider and astrophysics results in order to reveal the true nature of dark matter.

\section*{Acknowledgements}

I thank all colleagues who contributed to the realization of the LHC and the CMS experiment, and all colleagues who were involved in obtaining the search results explained here.  I also thank M. Mulders, H. Prosper and S. Worm for comments on the text.  I furthermore thank the organizers of the Balkan Workshop 2011, and acknowledge the travel support by the U.S Department of Energy grant No. DE-FG02-97ER41022.


\begin{thebibliography}{00}

\bibitem{LHC}
  L.~Evans, (ed.) and P.~Bryant, (ed.),
  ``LHC Machine,''
  JINST {\bf 3} (2008) S08001;
  http://lhc.web.cern.ch/lhc/

\bibitem{CMS}
  R.~Adolphi {\it et al.} [CMS Collaboration],
  ``The CMS experiment at the CERN LHC,''
  JINST {\bf 3 } (2008)  S08004.

\bibitem{alphaT}
  L.~Randall, D.~Tucker-Smith,
  ``Dijet Searches for Supersymmetry at the LHC,''
  Phys.\ Rev.\ Lett.\  {\bf 101 } (2008)  221803.
  [arXiv:0806.1049 [hep-ph]].

\bibitem{RA1}
  S.~Chatrchyan {\it et al.} [CMS Collaboration],
  ``Search for Supersymmetry at the LHC in Events with Jets and Missing Transverse Energy,''
  arXiv:1109.2352 [hep-ex], CMS-PAS-SUS-11-003.

\bibitem{razor}
  C.~Rogan,
  ``Kinematical variables towards new dynamics at the LHC,''
  [arXiv:1006.2727 [hep-ph]].

\bibitem{RAzr}
  S.~Chatrchyan {\it et al.} [CMS Collaboration],
  ``Search for supersymmetry with the razor variables at $\sqrt{s} = 7$~TeV,''
  CMS-PAS-SUS-11-008.

\bibitem{mt2}
  C.~G.~Lester and D.~J.~Summers,
  ``Measuring masses of semiinvisibly decaying particles pair produced at
  hadron colliders,''
  Phys.\ Lett.\  B {\bf 463} (1999) 99
  [arXiv:hep-ph/9906349], 
  A.~Barr, C.~Lester and P.~Stephens,
  ``m(T2) : The Truth behind the glamour,''
  J.\ Phys.\ G {\bf 29} (2003) 2343
  [arXiv:hep-ph/0304226].

\bibitem{RAMT2}
  S.~Chatrchyan {\it et al.} [CMS Collaboration],
  ``Search for supersymmetry in hadronic final states using $M_{T2}$ in 7 TeV pp collisions at the LHC,''
  CMS-PAS-SUS-11-005.

\bibitem{RA2}
  S.~Chatrchyan {\it et al.} [CMS Collaboration],
  ``Search for supersymmetry in all-hadronic events with missing energy,''
  CMS-PAS-SUS-11-004.

\bibitem{RA2b}
  S.~Chatrchyan {\it et al.} [CMS Collaboration],
  ``Search for New Physics in Events with b-quark Jets and Missing Transverse Energy in Proton-Proton Collisions at $\sqrt{7}$~TeV,''
  CMS-PAS-SUS-11-006.

\bibitem{RA3}
  S.~Chatrchyan {\it et al.} [CMS Collaboration],
  ``Search for Supersymmetry in Events with Photons, Jets and
Missing Energy,''
  CMS-PAS-SUS-11-009.

\bibitem{RA4}
  S.~Chatrchyan {\it et al.} [CMS Collaboration],
  ``Search for supersymmetry in pp collisions at $\sqrt{s} = 7$~TeV in events with a single lepton, jets, and missing transverse momentum,''
  CMS-PAS-SUS-11-015.

\bibitem{RA5}
  S.~Chatrchyan {\it et al.} [CMS Collaboration],
  ``Search for new physics with same-sign isolated dilepton events with jets and missing energy,''
  CMS-SUS-11-010.

\bibitem{RA6}
  S.~Chatrchyan {\it et al.} [CMS Collaboration],
  ``Search for new physics in events with opposite-sign dileptons and missing transverse energy,''
  CMS-PAS-SUS-11-011.

\bibitem{RAZ}
  S.~Chatrchyan {\it et al.} [CMS Collaboration],
  ``Search for New Physics in Events with a $Z$ Boson and Missing Transverse Energy'',
  CMS-PAS-SUS-11-017.

\bibitem{RAZj1}
  S.~Chatrchyan {\it et al.} [CMS Collaboration],
  ``Search for Physics Beyond the Standard Model in $Z + jets + \slashed{E}_T$ events at the LHC'',
  CMS-PAS-SUS-11-012.

\bibitem{RAZj2}
  S.~Chatrchyan {\it et al.} [CMS Collaboration],
  ``Search for Physics Beyond the Standard Model in $Z + jets + \slashed{E}_T$ events at the LHC'',
  CMS-PAS-SUS-11-019.

\bibitem{RA7}
  S.~Chatrchyan {\it et al.} [CMS Collaboration],
  ``Search for Physics Beyond the Standard Model Using Multilepton Signatures in pp Collisions at sqrt(s)=7 TeV,''
  [arXiv:1106.0933 [hep-ex]], CMS-SUS-10-008, CERN-PH-EP-2011-046.

\bibitem{RA22010}
  S.~Chatrchyan {\it et al.} [CMS Collaboration],
  ``Search for New Physics with Jets and Missing Transverse Momentum in pp collisions at sqrt(s) = 7 TeV,''
  [arXiv:1106.4503 [hep-ex]], CMS-PAS-SUS-10-005, CERN-PH-EP-2011-065.

\bibitem{cmshscp}
  V.~Khachatryan {\it et al.} [CMS Collaboration],
  ``Search for Heavy Stable Charged Particles in pp collisions at $\sqrt{s} = 7$~TeV,"
  CMS-EXO-11-022.

\bibitem{CRexcess}
  O.~Adriani {\it et al.}  [PAMELA Collaboration],
  ``An anomalous positron abundance in cosmic rays with energies 1.5-100 GeV,''
  Nature {\bf 458} (2009) 607
  [arXiv:0810.4995 [astro-ph]].

\bibitem{TeVDM}
  N.~Arkani-Hamed, D.~P.~Finkbeiner, T.~R.~Slatyer, N.~Weiner,
  ``A Theory of Dark Matter,''
  Phys.\ Rev.\  {\bf D79 } (2009)  015014.
  [arXiv:0810.0713 [hep-ph]].

\bibitem{cmsdarkphoton}
  S.~Chatrchyan {\it et al.} [CMS Collaboration],
  ``Search for Light Resonances Decaying into Pairs of Muons as a Signal of New Physics,''
  JHEP {\bf 1107 } (2011)  098.
  [arXiv:1106.2375 [hep-ex]], CMS-EXO-11-013, CERN-PH-EP-2011-064.


\end{thebibliography}
\end{document}